\documentclass[iop,apj]{emulateapj}
\usepackage{amsmath,amssymb,amstext,amsthm}
\usepackage{gensymb}
\usepackage{epsfig}
\usepackage{subfigure}
\usepackage{upgreek}
\usepackage{accents}
\usepackage{ulem} 

\usepackage[breaklinks,colorlinks,citecolor=blue,linkcolor=magenta]{hyperref} 

\usepackage[all]{hypcap} 

\usepackage{aas_macros}
\usepackage{natbib}
\usepackage{color}
\setcitestyle{notesep={ }}

\def\HI{{H{\small I}}}

\def\deg{{^{\circ}}}

\shorttitle{The 21-cm Global Signal}
\shortauthors{Burns et al.}

\begin{document}
\normalem

\title{A Space-Based Observational Strategy for Characterizing the First Stars and Galaxies Using the Redshifted 21-cm Global Spectrum}

\author{
 Jack~O.~Burns\altaffilmark{1},
 Richard Bradley\altaffilmark{6},
 Keith Tauscher\altaffilmark{1,2},
 Steven Furlanetto\altaffilmark{3},
 Jordan Mirocha\altaffilmark{3},
 Raul Monsalve\altaffilmark{1,5},
 David Rapetti\altaffilmark{1,9},
 William Purcell\altaffilmark{13},
 David Newell\altaffilmark{13},
 David Draper\altaffilmark{13},
 Robert MacDowall\altaffilmark{4},
 Judd Bowman\altaffilmark{5},
 Bang Nhan\altaffilmark{1,6},
 Edward~J.~Wollack\altaffilmark{4},
 Anastasia Fialkov\altaffilmark{8},
 Dayton Jones\altaffilmark{11},
 Justin C. Kasper\altaffilmark{7},
 Abraham Loeb\altaffilmark{8},
 Abhirup Datta\altaffilmark{1,12},
 Jonathan Pritchard\altaffilmark{10},
 Eric Switzer\altaffilmark{4},
 Michael Bicay\altaffilmark{9},
  }

\affil{$^{1}${Center for Astrophysics and Space Astronomy, Department of Astrophysical and Planetary Science, University of Colorado, Boulder, CO 80309, USA}}

\affil{$^{2}${Department of Physics, University of Colorado, Boulder, CO 80309, USA}}

\affil{$^{3}${Department of Physics and Astronomy, University of California at Los Angeles, Los Angeles, CA 90095, USA}}

\affil{$^{4}${NASA Goddard Space Flight Center, Greenbelt, MD 20771, USA}}

\affil{$^{5}${Arizona State University, School of Earth and Space Exploration, P.O. Box 876004, Tempe, AZ 85287, USA}}

\affil{$^{6}${National Radio Astronomy Observatory, 520 Edgement Road, Charlottesville, VA 22903, USA}}

\affil{$^{7}${Department of Climate and Space  Sciences and Engineering, University of Michigan, Ann Arbor, MI 48109, USA}}

\affil{$^{8}${Center for Astrophysics, 60 Garden St., MS 51, Cambridge, MA 02138, USA}}

\affil{$^{9}${NASA Ames Research Center, Moffett Field, CA 94035, USA}}

\affil{$^{10}${Department of Physics, Blackett Laboratory, Imperial College, London SW7 2AZ, UK}}

\affil{$^{11}${Space Science Institute, 4750 Walnut Street, Suite 205, Boulder, CO 80301, USA}}

\affil{$^{12}${Indian Institute of Technology, Indore, India}}

\affil{$^{13}${Ball Aerospace Corporation, 1600 Commerce Street, Boulder, CO 80301 USA}}

\email{E-mail: jack.burns@colorado.edu}
\begin{abstract}
{The redshifted 21-cm monopole is expected to be a powerful probe of the epoch of the first stars and galaxies ($10<z<35$).  The global 21-cm signal is sensitive to the thermal and ionization state of hydrogen gas and thus provides a tracer of sources of energetic photons -- primarily hot stars and accreting black holes -- which ionize and heat the high redshift intergalactic medium (IGM).  This paper presents a strategy for observations of the global spectrum with a realizable instrument placed in a low altitude lunar orbit, performing night-time 40-120 MHz spectral observations, while on the farside to avoid terrestrial radio frequency interference, ionospheric corruption, and solar radio emissions.  The  frequency structure, uniformity over large scales, and unpolarized state of the redshifted 21-cm spectrum are distinct from the spectrally featureless, spatially-varying, and polarized emission from the bright foregrounds. This allows a clean separation between the primordial signal and foregrounds. For signal extraction, we model the foreground, instrument, and 21-cm spectrum with eigenmodes calculated via Singular Value Decomposition analyses. Using a Markov Chain Monte Carlo algorithm to explore the parameter space defined by the coefficients associated with these modes, we illustrate how the spectrum can be measured and how astrophysical parameters (e.g. IGM properties, first star characteristics) can be constrained in the presence of foregrounds using the {\it Dark Ages Radio Explorer} (DARE).} 

\end{abstract}

\section{Introduction}
One of the last frontiers of observational cosmology is the time period stretching from the end of the Dark Ages through Cosmic Dawn ($\approx$80$-$500 million years after the Big Bang).  This is a virtually unobserved yet key epoch in the early Universe. During this interval, the first luminous objects including stars, galaxies, and accreting black holes ``turned on''  \citep[e.g.,][]{Loeb-Furlanetto2013}.  Furthermore, this time period saw the birth of structural complexity in the Universe.  At the beginning of the Dark Ages, corresponding to the Epoch of Recombination, the Universe was smooth to 1 part in $10^5$ as evidenced by the Cosmic Microwave Background \citep[CMB; e.g.,][]{2013pss6.book..609M}.  Yet less than a billion years later, the Universe was teeming with complex structure.  Thus, this transition time in the Universe is vital to understanding  how the core components and structures of today's Universe came to be.

The highly-redshifted 21-cm spectral line of neutral hydrogen, produced by a spin-flip hyperfine transition \citep{1951Natur.168..356E, 1958PIRE...46..240F}, provides an observable window into the early Universe's intergalactic medium (IGM) before the Epoch of Reionization (EoR) was complete \citep{madau97}.  The heating and ionization caused by the ``first objects to light up the Universe''\footnote{NRC Astrophysics Decadal Survey: New Worlds, New Horizons in Astronomy and Astrophysics, {http://www.nap.edu/catalog/12951/new-worlds-new-horizons-in-astronomy-and-astrophysics}.} serve as indirect probes of the nature of the first stars and galaxies.  With an effective optical depth of $\approx$1\% and sensitivity to low temperatures, the resulting signal measured against the CMB permits us to investigate a large evolutionary range from the Dark Ages through the end of the EoR \citep[e.g.,][]{2006PhR...433..181F, morales10, pritchard12}.

\begin{figure*}[t!]
\begin{center}
\includegraphics[width=0.7\textwidth]{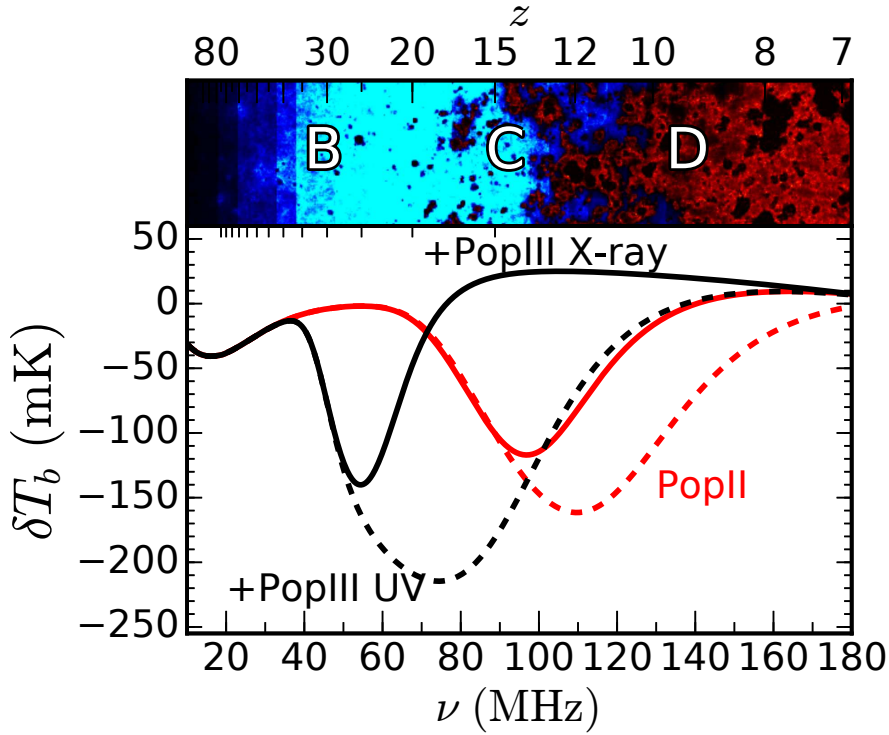}
\caption{Evolution of a slice of the Universe, from early times (left, upper panel) to late times (right) as well as several different models for the corresponding 21-cm spectrum relative to the CMB (lower panel). The red lines are conservative estimates with Pop II (metal-rich) stars only, while the black curves assume that Pop III (metal-free) star formation also occurs, but only in low-mass galaxies where atomic cooling is inefficient. The dashed and solid black curves assume that Pop III stars are distinct from Pop II stars in their emission properties -- 100 times brighter in the UV (dashed) and in the UV + X-ray emissions (solid), respectively. The dashed red curve assumes stellar properties corresponding to low redshift Pop II stars whereas the solid red curve corresponds to metallicities of 5\% solar.  Designations B, C and D indicate the redshift corresponding to the ignition of first stars, the formation of initial black hole accretion, and the onset of reionization, respectively. See Section 6 for further discussion. Figure adapted from \cite{2010PhRvD..82b3006P} using the new reference models from \cite{2017MNRAS.464.1365M}.}

\label{fig:global_signal}
\end{center}
\end{figure*}

The 21-cm all-sky or global signal \citep{1999A&A...345..380S, 2010PhRvD..82b3006P} (Figure \ref{fig:global_signal}) is an attractive observational target for either a single antenna \citep[e.g.,][]{2010Natur.468..796B, 2012AdSpR..49..433B, 2013ExA....36..319P,  2014ApJ...782L...9V, 2015ApJ...799...90B, 2015PASA...32....4S} or a small, compact array of antennas \citep[e.g.,][]{2015ApJ...809...18P, 2015MNRAS.450.2291V, 2015ApJ...815...88S}.  Features in the spectrum may provide the first constraints on the birth and nature of the first luminous objects \citep[e.g.,][]{ 2006MNRAS.371..867F}.  Such an experiment for 21-cm cosmology is analogous to the COBE measurement of the CMB blackbody spectrum, which set the stage for more detailed studies of spatial fluctuations by WMAP and Planck.  

In this paper, we describe {a space-based strategy for observations of the 21-cm global signal that probes} the time of formation and the characteristics of the first stars and galaxies.  We demonstrate how signal extraction using a realizable radiometer system and Bayesian statistical techniques, in the presence of strong galactic and extragalactic foregrounds, can measure spectral features and constrain the physical properties of the first luminous objects. {We use the new detailed design of the \emph{Dark Ages Radio Explorer} (DARE) to illustrate how the 21-cm spectrum can be extracted from the foreground using a feasible observational strategy.}  

DARE is proposed to conduct observations between 40 and 120 MHz in an orbit around the Moon with data taken only above the lunar farside.  On Earth, the ionosphere corrupts low frequency observations \citep[see e.g.,] [and references therein]{2014MNRAS.437.1056V, 2015MNRAS.453..925V, 2015RaSc...50..130R, 2015ApJ...813...18S, 2016ApJ...831....6D} due to refraction, absorption, and emission driven by solar emissions and the solar wind \citep{1990ibs..symp..153D, 2011JGRA..116.9307L, 2003MNRAS.343..725P}.  At 50 km above the lunar farside, $>$90 dB of radio frequency interference (RFI) attenuation produces an environment quiet to $<$1 mK \citep[e.g.,][]{2013AJ....145...23M}. In addition, the Moon shields the instrument (about half the time) from variable solar emission caused by flares and coronal activity \citep[e.g.,][]{1997ApJ...474L..65M}.  Therefore, observations above the night-time, pristine, radio-quiet lunar farside (as verified by RAE-2, \citealt{1976JGR....81.5948A}) bypass the challenges presented by the Earth and the Sun and provide an optimal site for measurements of the global 21-cm signal.

The key insight permitting the Cosmic Dawn signal to be detected in the presence of bright foregrounds is that once the Moon blocks solar effects and terrestrial RFI, the foregrounds are significantly different in their characteristics from the expected 21-cm spectral signal.  The 21-cm monopole strength is about four orders of magnitude weaker than the Galactic foreground. However, the  21-cm signal is separable from the foreground because it is spatially uniform at angular scales $\gtrsim$10$\degree$ \citep[e.g.,][]{2011JCAP...04..038B, 2013PhRvD..87d3002L}, unpolarized, and has distinct spectral features whereas the observed foreground varies spatially, exhibits polarized emission, and is spectrally featureless.  The 21-cm cosmological signal can then be extracted using algorithms similar to those employed for CMB observations implemented via a Markov Chain Monte Carlo framework \citep{2012MNRAS.419.1070H, 2016MNRAS.455.3829H}.

The paper is organized as follows.  In Section \ref{sec:strategy}, we introduce and summarize the space-based observational strategy. In Section \ref{sec:theory}, an overview of the stellar models for the sky-averaged 21-cm signal used to develop the observational strategy is presented. Section \ref{sec:foregrounds} describes the nature and brightness of astronomical foregrounds which must be considered in efforts to measure the much weaker Cosmic Dawn signal. Section \ref{sec:instrument} provides a synopsis of the new design for DARE. Section \ref{sec:pipeline} describes our software pipeline for signal extraction. In Section \ref{sec:estimates}, we discuss the physical parameters (and their uncertainties) associated with the first stars, black holes, and galaxies that are expected to be measured using the 21-cm all-sky spectrum.  Section \ref{sec:conclusion} presents a summary of the potential use of the 21-cm background to detect the first luminous objects in the early Universe.

\section{Summary of the Observational Strategy}
\label{sec:strategy}

{Here we briefly describe key aspects of our observational strategy. The following sections will provide details about each item, as well as their relevance to the overall tactic. The core components of the strategy are as follows:
\begin{itemize}
\item We incorporate a wide range of theoretical models ($> 1.5\times10^4$) from two different classes of possible signals, differing by the generation of stars whose contribution dominates the behavior of the signal (Pop II or Pop III). We show that the DARE instrument in orbit of the Moon can effectively differentiate between these models using our Bayesian inference pipeline.
\item We realistically model the diffuse foregrounds accounting for spatial variations of their spectral index, which is estimated from all-sky, publicly available maps at two frequencies ($45$ and $408$ MHz). 
\item The new DARE reference instrument design incorporates (1) an optimized antenna with on-orbit beam calibration, (2) the replacement of Dicke switches for bandpass calibration with a pilot frequency tone system capable of high dynamic range monitoring of gain variations and measurements of the system reflection coefficients, and (3) polarimetric observations to provide a model-independent measure of the beam-averaged foregrounds. The observations, performed from the radio-quiet zone above the Moon's farside, will be enabled through a unique ``frozen'' $50\times125$ km lunar equatorial orbit \citep{2017arXiv170200286P}. The nominal observation time corresponds to $800$ hours, which results in radiometric noise integration to the 1.7 mK level at 60 MHz. The instrument provides data with the frequency range ($40-120$ MHz), spectral resolution (50 kHz), beam characteristics (${\approx}60\deg$ FWHM at 60 MHz), and polarization required to measure the spectral features expected from the wide range of theoretical models considered.   DARE's present-day observing strategy utilizes four quiet-sky pointing directions away from the galactic center.
\item Our signal extraction pipeline is centered around a Singular Value Decomposition (SVD) approach, which allows us to robustly separate the $21$-cm signal from the additional contributions to the measurement by using orthogonal modes of variation of each component. These modes are determined from well-characterized training sets constructed from either theory or measurements.
\end{itemize}
We constructed a detailed end-to-end observation model that generates simulated antenna temperatures using our models for the diffuse foregrounds and the 21-cm spectrum, and the predicted telescope pointing, Moon location, and instrument characteristics.  Unlike previous papers (e.g., \citealt{2015ApJ...813...11M, 2016MNRAS.455.3829H}), which assumed perfect knowledge of the instrument, this new process accounts for and propagates the uncertainty in the instrumental parameters to the signal extraction pipeline. Our instrument sensitivity metric is defined as the RMS uncertainty of the extracted $21$-cm spectrum, averaged over the observation band. Our requirement for this metric is to keep it below $20$ mK for all models processed with our pipeline.}

\section{Models for the 21-cm Global Signal}
\label{sec:theory}
{In this section we discuss the global $21$ cm signal, and describe the broad set of physical models that are incorporated into our analysis strategy.}

The 21-cm global signal arises from the radiation effects produced by the first stars, accreting black holes, and galaxies on the surrounding IGM. X-ray and UV emission from these objects and their descendants heated and ionized the tenuous gas that lies between galaxies, culminating in the Epoch of Reionization several hundred Myrs later. The 21-cm background can be used to measure these radiation effects with the hyperfine line of the neutral hydrogen (\HI) gas pervading the Universe. The expansion of the Universe redshifts these photons from earlier epochs to lower observed frequencies, $\nu$ =$1420/(1+z)$ MHz (e.g., at $z=30$, $\nu$ = 45 MHz).  Importantly, this frequency-redshift relation enables a direct reconstruction of the history of the Universe as a function of time from the 21-cm spectrum.   

Figure \ref{fig:global_signal} shows some example predictions (amongst those currently allowed) for the 21-cm spectrum during the Dark Ages and Cosmic Dawn.  The brightness temperature of this 21-cm signal is given by \citep[e.g.,][]{1997ApJ...475..429M, 1999A&A...345..380S, 2006PhR...433..181F}

\begin{equation}
\begin{split}
\delta{T_{b}} \simeq 27x_{\textsc{HI}}\left(\frac{T_s - T_{\mathrm{CMB}}}
{T_s}\right) \left(\frac{1+z}{10}\right)^{1/2}  \\
(1+\delta_B)  \left[\frac{\partial_r v_r}{(1+z)H(z)}\right]^{-1} \mathrm{mK}\,, 
\end{split}  
\label{eq:sig}
\end{equation}
where $x_{\textsc{HI}}$ is the fraction of neutral gas, $T_s$ is the 21-cm spin temperature, $T_{\mathrm{CMB}}$ is the CMB temperature, $\delta_B$ is the baryon overdensity (taken here to be $\delta_B \sim 0$), and $H(z)$ is the Hubble parameter. The last term in this equation includes the effect of the peculiar velocities with line of sight velocity derivative $\partial_r v_r$. {Since we will measure the spatially averaged ${\delta}T_b$, the effects of the last term in Equation \ref{eq:sig} are negligible for observations of the 21-cm global signal} \citep[e.g.,][]{2004MNRAS.352..142B, 2005ApJ...624L..65B}.

Several important physical processes drive the evolution of $\delta T_b$ with redshift. These include: (1) UV radiation from the first stars, which ``activates'' the spin-flip signal through the Wouthuysen-Field mechanism \citep{1952AJ.....57R..31W, 1958PIRE...46..240F}; (2) X-ray heating, likely generated by gas accretion onto the first black holes;  and (3) ionizing photons from the first galaxies (which destroy the neutral hydrogen).

The relevant radiation backgrounds grow at different times, so their interplay creates distinct features in the spectrum \citep{2006PhR...433..181F, 2010PhRvD..82b3006P, 2011MNRAS.411..955M}. When the first stars appear, their UV radiation drives $T_s$ toward the cold temperatures that are characteristic of IGM gas ($z\sim35-22$ across our range of models; Region B in Figure \ref{fig:global_signal}), triggering a deep absorption trough \citep{madau97}. Shortly after, black holes likely formed, e.g., as remnants of the first stars ($z\sim25-12$ across our range of models; Region C). The energetic X-ray photons from these accreting black holes travel great distances, eventually ionize H and He atoms, and produce photo-electrons that deposit some of their energy as heat in the IGM \citep{shull85,furl10-xray}, transforming the 21-cm signal from absorption into emission as the gas becomes hotter than the CMB (Region D). The emission peaks as photons from these stars and black holes ionize the IGM gas ($z < 12$), eventually eliminating the spin-flip signal.

\begin{figure*}[t!]
\begin{center}
\includegraphics[width=0.7\textwidth]{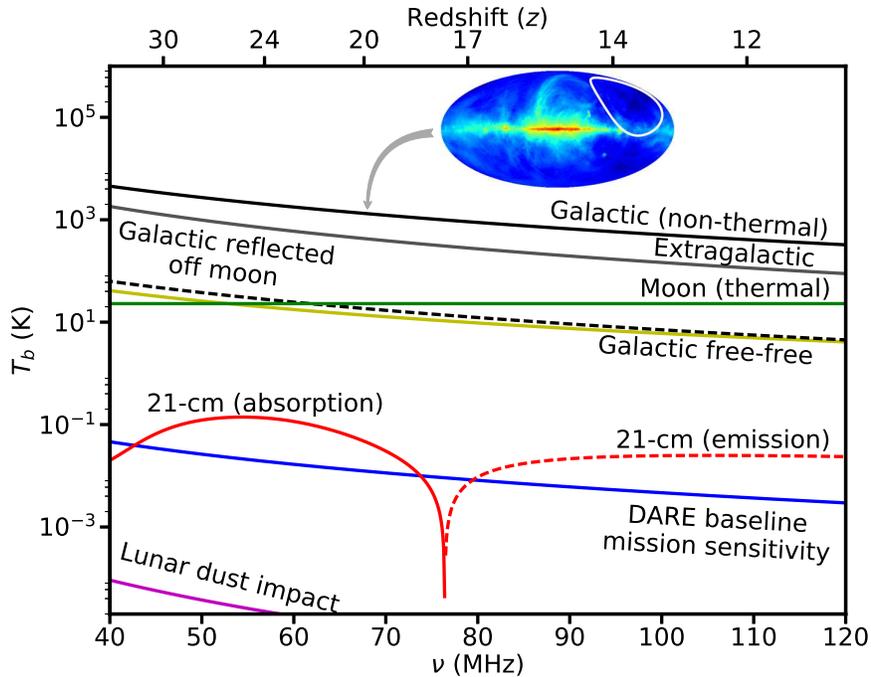}
\caption{The Galactic and Extragalactic spectra for a typical region away from the Galactic center. The Galaxy spectrum also “reflects” off the Moon \citep{1969ARA&A...7..201E}. The Moon's thermal emission at low radio frequencies arises from cold, uniform subsurface layers. The effects of hyperkinetic dust impacts on the spacecraft in orbit of the Moon are unimportant. The red curve illustrates the spectral features in the 21-cm spectrum, where the dashed part of the curve corresponds to emission and the solid to absorption for this log-linear plot.  {\it{Inset}}: A Mollweide projection of the sky at 408 MHz \citep{1982A&AS...47....1H} along with a DARE beam FWHM white contour. }

\label{fig:Foregrounds}
\end{center}
\end{figure*}

The dashed red curve in Figure \ref{fig:global_signal} assumes that the efficiency and properties of star formation in early galaxy populations  \citep{2016MNRAS.460..417S, 2017MNRAS.464.1365M}  and the relationship between X-ray luminosity and star formation rate are the same as at later times \citep{Mineo2012b}. There are several reasons to expect that this Pop II model is conservative{, i.e., that it underestimates the total production rate of UV and X-ray photons}. For example, it assumes solar metallicity, though stars in high-$z$ galaxies are likely forming in metal-poor environments, which can boost their UV \citep{2009MNRAS.400.1019E} and X-ray outputs \citep{2016MNRAS.457.4081B}. The solid red curve in Figure \ref{fig:global_signal} assumes that galaxies have metallicities ($Z$) 5\% of solar, which results in a shallower absorption feature due to enhanced X-ray emission \citep[assuming the][$L_X$-SFR-$Z$ relation]{2016MNRAS.457.4081B}. Alternatively, the black curves include a simple model for Pop III stars, in which low-mass halos (below atomic cooling threshold) can produce UV and X-ray photons (neglected by red curves). Boosts of 100 in the efficiency of the UV (dashed black) and also the X-ray luminosity (solid black) of Pop III stars relative to Pop II result in a variety of qualitatively different predictions for the global 21-cm signal. Pop III models that resemble our black curves should be relatively straightforward to distinguish from Pop II-only models for an experiment like DARE (see Section 6 and Figure \ref{fig:finalspectra}). At this stage, our ability to label each set of curves as being Pop II- or Pop III-dominated assumes that the current Pop II models are well calibrated \citep{2017MNRAS.464.1365M}. New measurements (by e.g., JWST) in the coming years can be immediately incorporated into the model, and will act to mitigate degeneracies between Pop II and Pop III sources. More subtle features of the signal, such as its asymmetry, may also reveal the presence of Pop III despite uncertainties in the calibration of Pop II models (Mirocha et al., in preparation).

It is also worth noting that the 21-cm global signal traces the collective effects of all sources in the redshift ranges illustrated in Figure \ref{fig:global_signal}, which form a mostly unresolved sea of fainter objects that likely dominate the total emissivity of the early Universe.  The red curves in Figure \ref{fig:global_signal} are calibrated to match the latest luminosity function measurements from HST (which probe relatively bright galaxies that can be resolved) and CMB optical depth ($\tau_e$) measurements from Planck \citep{2017MNRAS.464.1365M}, with variations arising solely due to differences in the adopted properties of galaxies beyond the current detection threshold.  JWST and future CMB missions will further constrain the bright-end of the galaxy luminosity function and $\tau_e$, respectively, and will thus enhance the sensitivity of the 21-cm global signal to Pop III stars and their remnants in faint galaxies.

{The signal models described in this section are used to create the signal training set, an essential component of our observational strategy from which the signal extraction pipeline calculates the main modes of variation of the signal. The new data expected from JWST and from CMB missions will constrain parameter space, which will allow us to restrict the training set and reduce parameter degeneracies and covariances. See Section~\ref{sec:pipeline} for more details on the training set and its effect on the uncertainty of our signal estimate.}

\section{Foregrounds}
\label{sec:foregrounds}

{Here, we discuss the origin and properties of the foregrounds expected in the $21$-cm measurement from lunar orbit, which are modeled and accounted for in our signal extraction pipeline.}

\subsection{Galaxy/Extragalactic Foregrounds}
Beam-averaged diffuse sky foregrounds represent the strongest contributors to any highly redshifted 21-cm measurement for a space-based experiment.  The most important arises from our Galaxy \citep{1999A&A...345..380S}.   In addition, a ``sea'' of Extragalactic sources appear as another diffuse, spectrally-featureless power-law foreground (at DARE's resolution) and contributes $\sim$10\% of the total sky brightness temperature (Figure \ref{fig:Foregrounds}). The emission from these foregrounds is produced by synchrotron radiation that intrinsically has a smooth frequency spectrum \citep[e.g.,][]{2015ApJ...799...90B, 2011MNRAS.413.2103P}. {On top of the spectral smoothness, the foregrounds are spatially variable (inset in Figure \ref{fig:Foregrounds}). Their featureless spectrum and spatial variability contrast with the spectral features and spatial uniformity of the 21-cm spectrum, making them separable \citep{2013PhRvD..87d3002L, 2014ApJ...793..102S}.}

Theoretical models predict that the foreground is well approximated by a third-order polynomial to levels below the amplitude of the 21-cm global signal, especially over low-foreground regions \citep{2015ApJ...799...90B}. Smoothness over a frequency range much broader than $\sim$40-120 MHz is supported by sky models produced from measurements that cover the range 10 MHz - 5 THz \citep{2008MNRAS.388..247D, 2017MNRAS.464.3486Z, 2017AJ....153...26S}.  These models rely on, at most, five components to describe the spectral content of the foreground over several decades in frequency. Global measurements from the Experiment to Detect the Global EoR Signature (EDGES), Sonda  Cosmol\'ogica  de  las  Islas  para  la  Detecci\'on  de Hidr\'ogeno Neutro (SCI-HI), Shaped Antenna measurement of the background RAdio Spectrum (SARAS), and Large-Aperture Experiment to Detect the Dark Ages (LEDA) provide further validation of the intrinsic foreground smoothness \citep{2008AJ....136..641R, 2017MNRAS.464.4995M, 2014ApJ...782L...9V, 2015ApJ...801..138P, 2016MNRAS.461.2847B}.

{In our strategy, we use a diffuse foreground model produced from all-sky observations taken at two frequencies, $45$ and $408$ MHz~\citep{1982A&AS...47....1H, 2011A&A...525A.138G}, in order to account for spatial variations in the spectral index.}

The spectrally smooth foreground is altered via the frequency-dependent antenna response \citep{2014MNRAS.437.1056V, 2015ApJ...799...90B, 2016MNRAS.455.3890M}. The beam directivity of finite-sized, wideband antennas does not remain constant across frequency \citep{Rumsey}. This beam ``chromaticity'' impacts the observed spectrum of the spatially-dependent foregrounds. The variation with frequency of the beam shape and directivity imprints spectral structure into the beam-averaged response that is not intrinsic to the foregrounds. 

As part of our strategy, chromaticity is addressed by minimizing instrumental design effects and making precise beam measurements on the ground and on-orbit. {We also estimate the beam chromaticity by modulating the beam-averaged foregrounds through rotation of the antenna about the boresight axis. This technique is discussed in Section \ref{sec:instrument}. This represents a significant advancement over previous simulations. For instance, in \cite{2016MNRAS.455.3829H}, the beam was taken to be Gaussian and the integrated foreground was assumed to perfectly follow a polynomial of the form $\ln{(T)}=\sum_{i=0}^5a_i\ln{(\nu)}^i$.}

{Finally, we note that the low foreground areas of the sky are polarized to a few percent ($\lesssim 5\%$) \citep{2014A&A...568A.101J, 2015A&A...583A.137J, 2016ApJ...830...38L}. Our dual polarization instrument directly measures this intrinsic sky polarization. At the same time, this polarization is minimized through dilution produced by our wide antenna beam, and also averaged down by our scanning strategy, which includes antenna rotation.}

\subsection{Other Foregrounds}

21-cm cosmology experiments in lunar orbit will also detect emission from the Moon via the antenna backlobe. The lunar spectrum is comprised of (1) thermal emission from a $\sim$100 m subsurface layer (i.e., electrical skin depth of the regolith) \citep{1971JAnSc..18..236S, 1975Icar...24..211K}  and (2) reflected Galactic emission, requiring a parameter in the data analysis pipeline to describe the Moon's reflectivity \citep{1964JGR....69.3257D, 2015MNRAS.450.2291V}.  

Other processes have a minor effect on the spectrum. Hyperkinetic impacts of dust from the interplanetary medium and the lunar exosphere on the spacecraft surface generate radio transients \citep[e.g.,][]{1985AdSpR...5...37M}; but the dust distribution around the Moon \citep[e.g.,][]{2010P&SS...58..830S}, the capacitance of the spacecraft, and solar wind conditions produce most of its emission at frequencies $<40$ MHz \citep[Figure \ref{fig:Foregrounds};][]{2013SoPh..286..549L}.

Bright, transient, nonthermal emission from Jupiter and Io also occur at $<$40 MHz \citep{2013P&SS...77....3P, 2012P&SS...61...32C}; however, at 40-120 MHz, the antenna temperature observed by an instrument like that proposed for DARE is only $\sim$1 mK for Jupiter \citep{2004ASPC..321..160Z}. Jupiter, and other  astronomical sources such as Cas A (similarly beam-diluted), may introduce low-level spectral effects due to scattering off the spacecraft.  Electromagnetic analysis, incorporating accurate models of the spacecraft, must be used to assess and calibrate these effects as part of the signal extraction pipeline.

Finally, atoms (e.g., carbon) in cold, diffuse gas in the Milky Way (and possibly in the IGM) produce radio recombination lines \citep[RRLs;][]{2011A&A...525A.128P, 2014ApJ...795L..33M}. These lines are sharp ($\sim$10 kHz wide), but spaced at known intervals of $\sim$1 MHz. Spectral channels containing RRLs constitute a negligible fraction of the data and may be discarded. Removal of potential RRLs from the 21-cm spectrum will drive the spectral resolution of the science instrument.  Also, beam dilution is expected to significantly reduce any impact from recombination lines.

\begin{figure*}[t!]
\begin{center}
\includegraphics[width=0.5\textwidth]{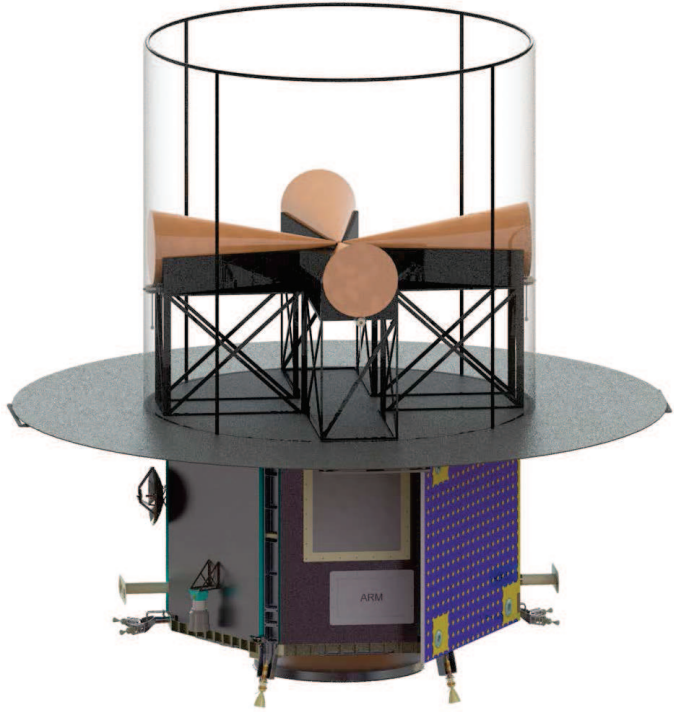}
\caption{An artistic rendering of the DARE observatory.  The science instrument thermal shield surrounds the antenna (shown transparent for clarity).  The antenna consists of a pair of dual, crossed bicones.  Beneath the antenna support structure is a deployed ground plane which aids in shaping the beam directivity.  Below the instrument is the spacecraft bus including the solar panels and telemetry system.}
\label{fig:Spacecraft}
\end{center}
\end{figure*}

\section{The DARE Science Instrument}
\label{sec:instrument}

To illustrate how the cosmological 21-cm spectrum can be extracted from the foregrounds, we use the {new} science instrument proposed for DARE (Figure \ref{fig:Spacecraft}). {In   \cite{2012AdSpR..49..433B}, we outlined a very basic approach for idealized, lunar-based 21-cm cosmology observations. We have now advanced the fidelity of the instrument model to evaluate measurements of the spectrum and constrain parameters for the first luminous objects at the significance level presented in Section \ref{sec:strategy}, in the presence of realistic and well modeled uncertainties.}

\begin{figure*}[t!]
\begin{center}
\includegraphics[width=0.5\textwidth]{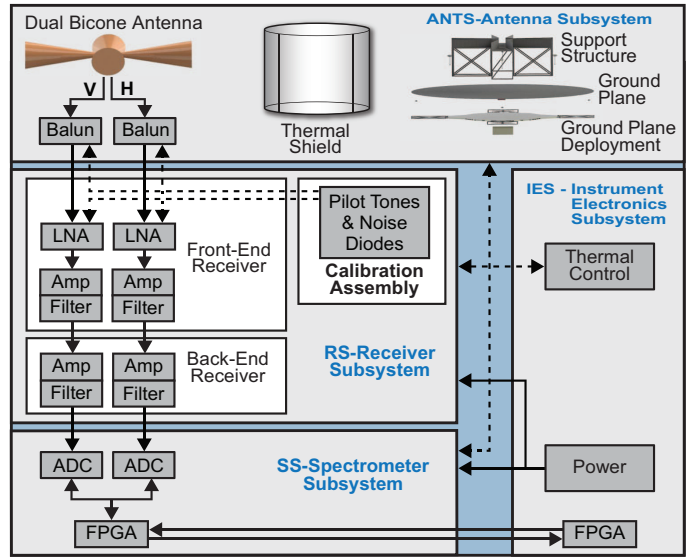}
\caption{DARE instrument block diagram.  DARE consists of four subsystems: dual polarization antenna, pilot tone calibration receiver, high resolution digital spectrometer, and a standard instrument electronics module for power, data handling, and instrument control that interfaces with the spacecraft.}
\label{fig:Instrument}
\end{center}
\end{figure*}

Figure \ref{fig:Instrument} shows a block diagram of the current DARE instrument design, which consists of four subsystems: (1) an antenna composed of a pair of crossed biconical dipoles above a ground plane that provides dual polarization with low reflection coefficient (-12 dB average across the band) and beam chromaticity (the beam directivity spectral knowledge goal is $\sim$20 ppm, see below); (2) {a thermally-controlled receiver with a calibration architecture that utilizes precise continuous-wave frequency tones optimized to yield a frequency response that meets DARE's RMS criterion}; (3) a spectrometer with a wide bandwidth and digital receiver that provides the spectral resolution and Stokes processing of the V and H channels; and (4) an instrument electronics subsystem to interface with the spacecraft. The expectation for the instrument envelopes the hardware performance of systems on the ground (e.g., EDGES, \citealt{2010Natur.468..796B}; Cosmic Twilight Polarimeter, CTP, \citealt{2017ApJ...836...90N}) and in space (Global Precipitation Measurement Microwave Imager, GMI\footnote{https://pmm.nasa.gov/gpm/flight-project/gmi}).

We model the forward instrument response following that used by EDGES \citep{2017ApJ...835...49M} as:

\begin{equation}
P =  g\left [|F|^2(\eta_lT_A+(1-\eta_l)T_{Ap})(1-|\Gamma_A|^2)+T_{\it off} \right ], 
\label{eq:instresponse}
\end{equation}

\noindent where $P$ is the raw power measured by the instrument, $g$ and $T_{\it off}$ represent the system gain and radiometric offset respectively, $\eta_l$   accounts for the antenna and balun losses at physical temperature $T_{Ap}$, $1-|\Gamma_A|^2$  accounts for the reflection coefficient of the antenna, and $|F|^2$  is the throughput of the receiver front end accounting for multiple reflections between the receiver and antenna.  The instrument calibration activities consist of using ground, on-board and on-orbit calibration to invert the forward instrument response model and provide an estimate of the antenna temperature $T_A$.

During science observations, the receiver is calibrated continuously using the pilot tone injection receiver architecture. The calibration system generates tones at $\approx$5 frequencies simultaneously across the band to adequately sample the frequency range.  The tones are each within a single 50 kHz spectrometer bin, and thus produce negligible degradation in spectral performance. The nominal calibration cycle consists of a sequence of four states which are enabled for 10 seconds each: 1) high-level tones directed toward the receiver, 2) low-level tones directed toward the receiver, 3) high-level tones directed toward the antenna, and 4) low-level tones directed toward the antenna. The gain of the receiver is computed by differencing instrument-measured power from the high- and low-level injected tones toward the receiver divided by the difference in effective input brightness temperature of the tones characterized during pre-flight calibration.  Likewise, the tones injected toward the antenna afford an on-board measurement of the antenna reflection computed in a similar fashion.

The terms $T_{\it{off}}$, $\eta_l$, and $|F|^2$  in Equation \ref{eq:instresponse} are computed based on ground measurements and the on-orbit trending of the receiver gain and reflection coefficient.

The DARE antenna and receiver are designed to minimize temperature variations by limiting exposure to the solar flux and lunar albedo.  For the antenna, thermal baffles, as shown in Figure \ref{fig:Spacecraft}, result in a predicted physical temperature change over the lunar orbit of $10\deg$C allowing DARE to maintain a nearly constant beam directivity.  The front-end receiver includes a proportional-integral-derivative temperature control to provide predicted thermal stability of $0.1\deg$C, thus reducing receiver systematics to meet DARE's calibration and stability requirements.

A novel feature of the current design is on-orbit calibration of the antenna directivity. Measurements of the beam will be obtained by receiving $\geq$3 narrow-band, circularly-polarized signals spaced across the band, transmitted from a large antenna on the Earth as DARE orbits the Moon above the nearside. The spacecraft (and antenna) is slowly rotated while it continuously receives these signals.  The received signal power at each frequency as a function of antenna pointing will produce a slice through the beam power pattern.  The transmitted signals will also reflect off the lunar regolith and return to the same antenna on Earth to correct for ionospheric effects.  The in-situ beam measurement system is currently being baselined to use the 140-foot radio telescope at the Green Bank Observatory, operating with $50\%$ aperture efficiency, transmitting 10 kW of power, and using 10 second averaging.  

Another innovation in the current design of DARE is polarization measurements to constrain and distinguish the beam-averaged foregrounds from the unpolarized {\HI}  signal \citep{2017ApJ...836...90N}. Our observation strategy incorporates rotation of the antenna about the boresight axis to modulate the signals captured by the two polarization arms of the antenna. This modulation results in induced polarization that tracks and measures the beam-averaged foreground spectrum, without relying on e.g., polynomial model fits, and is insensitive to the spatially uniform 21-cm signal.  CMB polarization measurements use analogous modulation approaches, achieving stability and systematic control required for ${\mu}K$ polarimetric sensitivity  \citep[e.g.,][]{2003ApJ...583....1B, 2013ApJ...768....9B}. 

{The on-orbit measurements of the antenna directivity and the induced polarization technique enable us to anticipate a knowledge of the beam-averaged foregrounds at a level of $\sim$20 ppm. This represents an important advancement that allows us to achieve our goal of $<$20 mK RMS spectral uncertainty on the extracted $21$-cm models.}

\section{Extracting the Cosmic 21-cm Spectrum}
\label{sec:pipeline}

\begin{figure*}[t!]
\begin{center}
\includegraphics[width=0.33\textwidth]{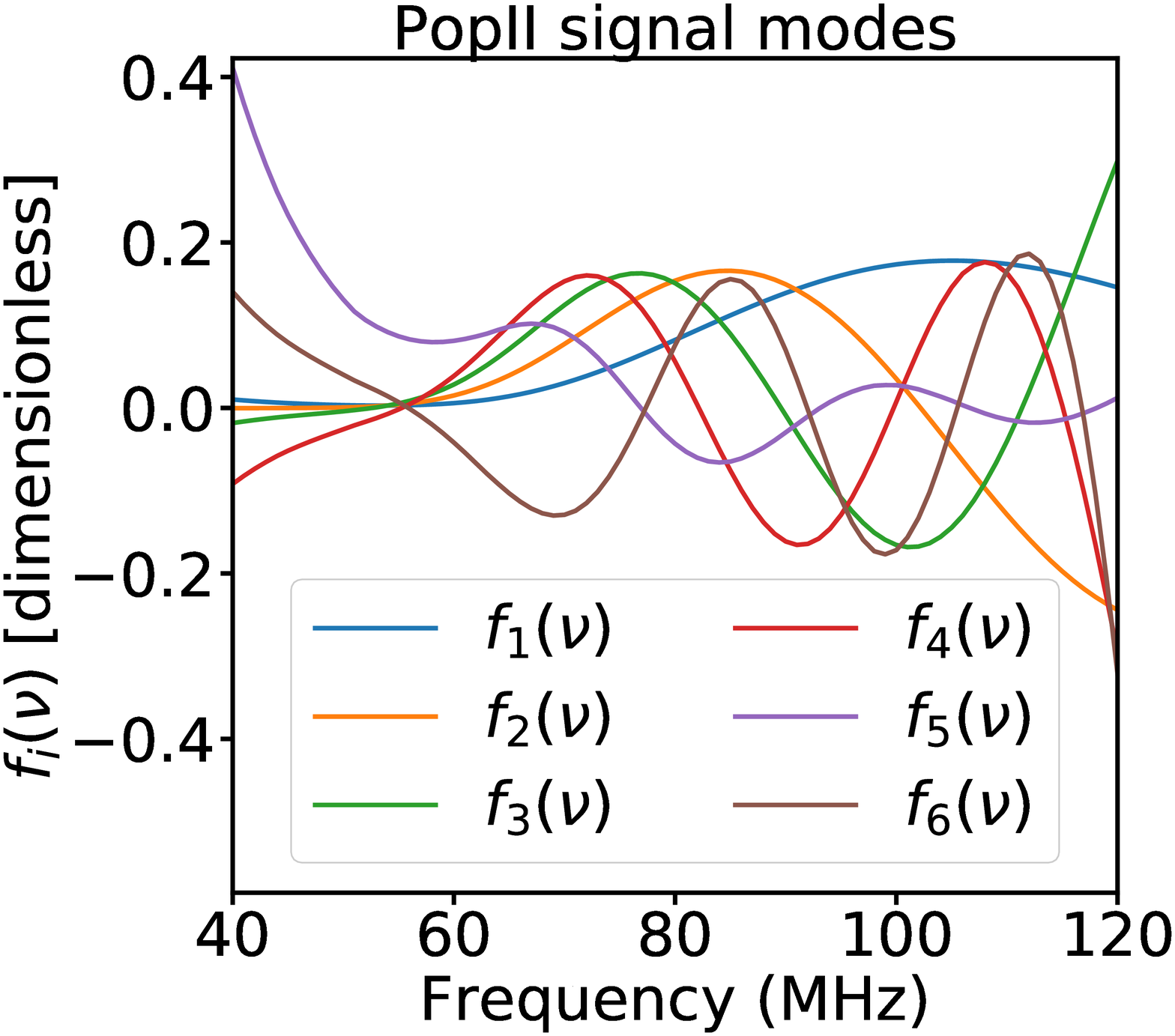}
\includegraphics[width=0.33\textwidth]{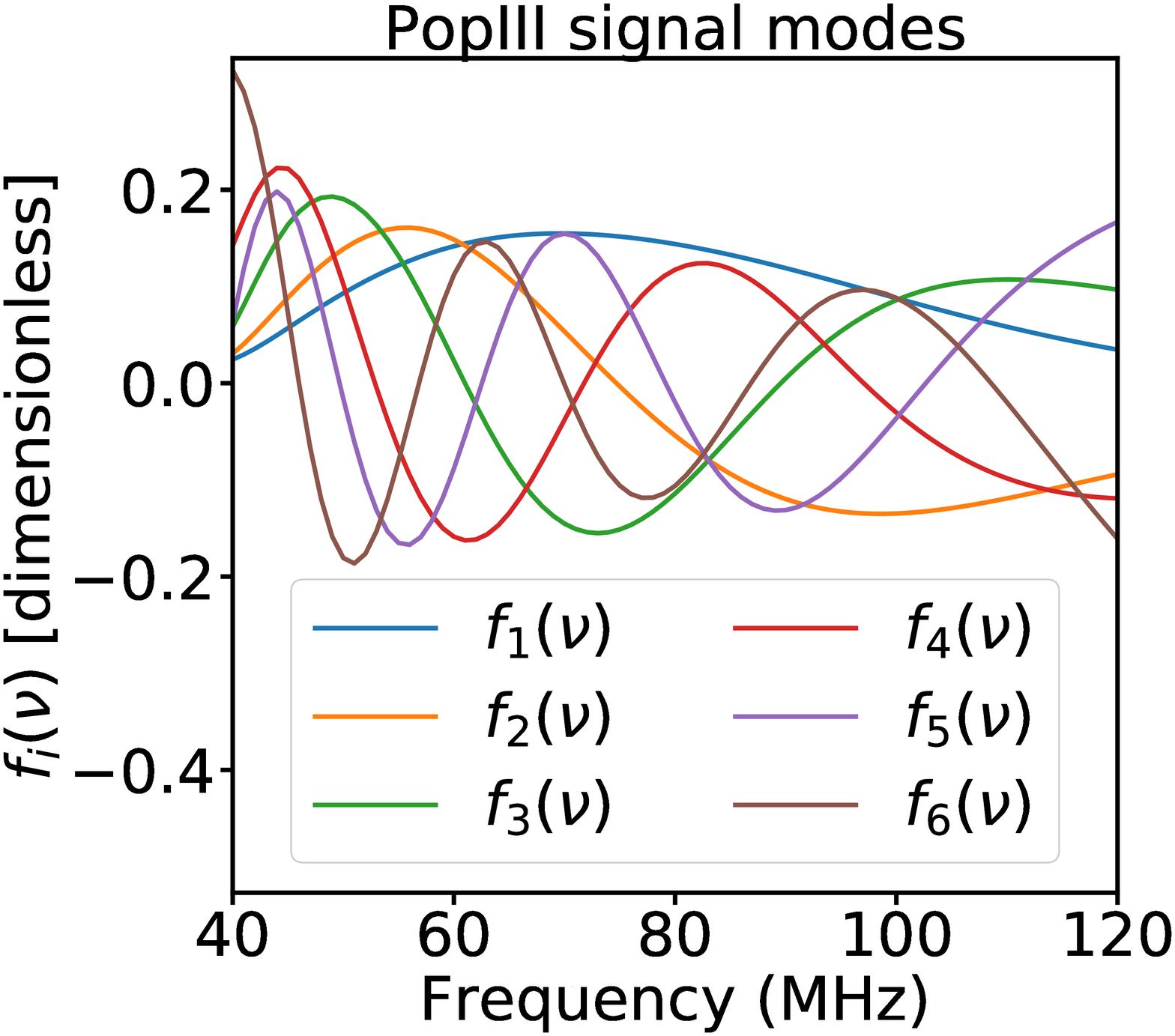}
\includegraphics[width=0.33\textwidth]{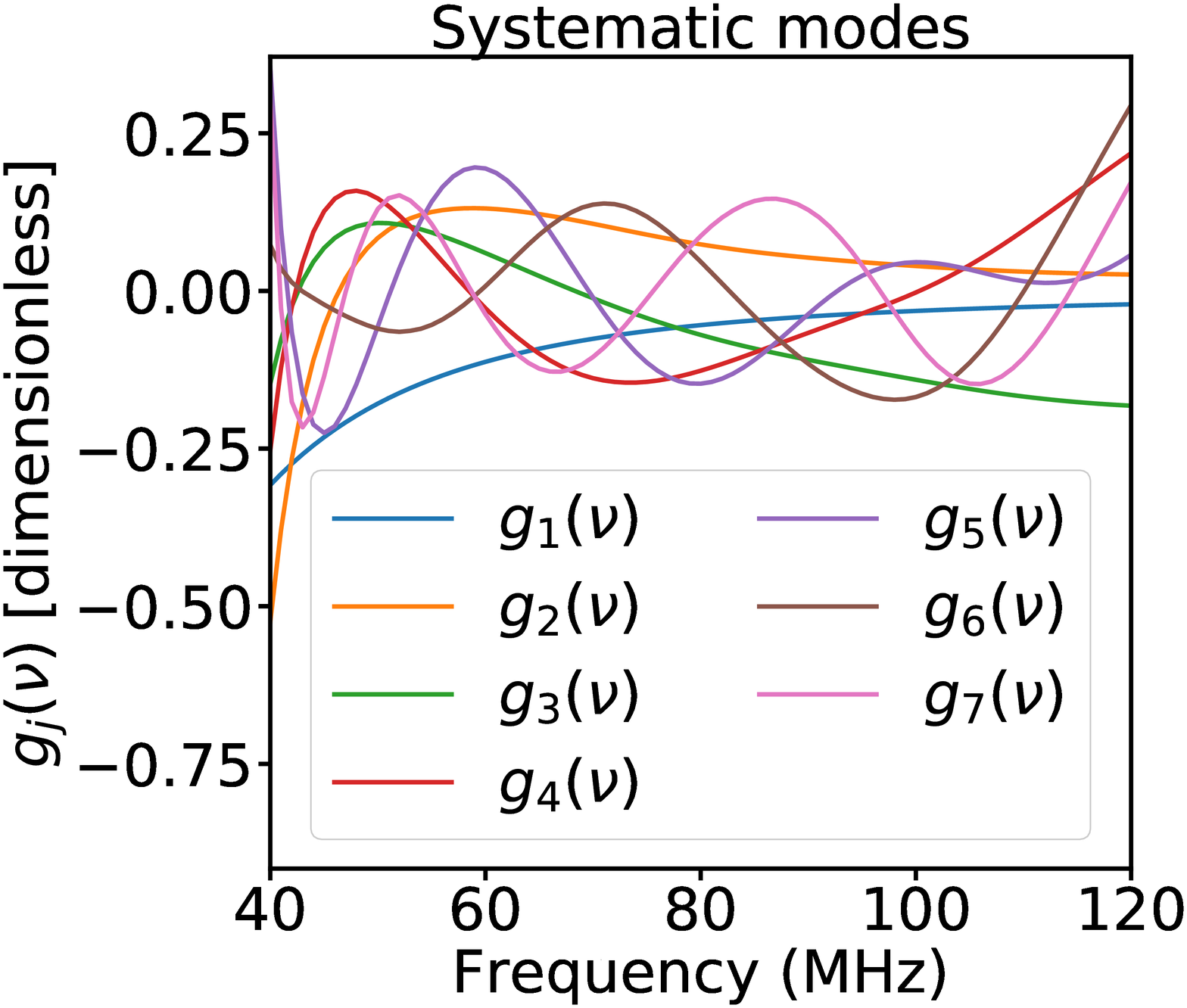}
\caption{{Left panel: the 6 principal SVD signal modes derived from 21-cm spectrum simulations of models based on primordial Pop II stars. Middle panel: the same but for signal models based on primordial Pop III stars. Right panel: the 7 principal SVD systematic modes derived from simulations of the instrument plus foreground. Each panel contains a set of orthonormal models, i.e. the curves represent only dimensionless shapes which are then multiplied by coefficients with units of temperature (K). The ability to separate the 21-cm signal from DARE's systematics hinges on the ability to distinguish between the signal modes, $f_i(\nu)$, and the systematic modes, $g_j(\nu)$.}}
\label{fig:modes}
\end{center}
\end{figure*}

\begin{figure*}[t!]
\begin{center}
\includegraphics[width=0.49\textwidth]{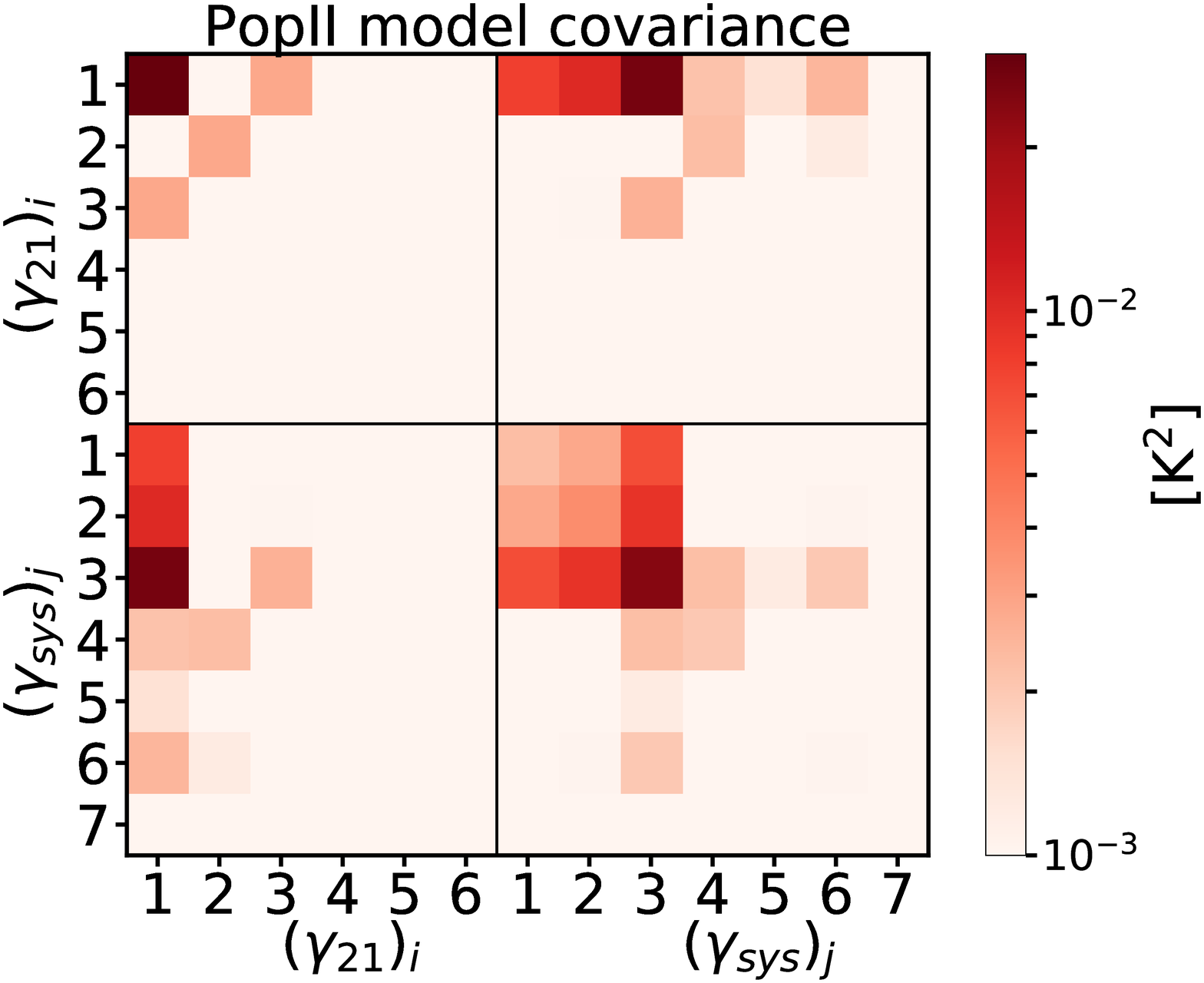}
\includegraphics[width=0.49\textwidth]{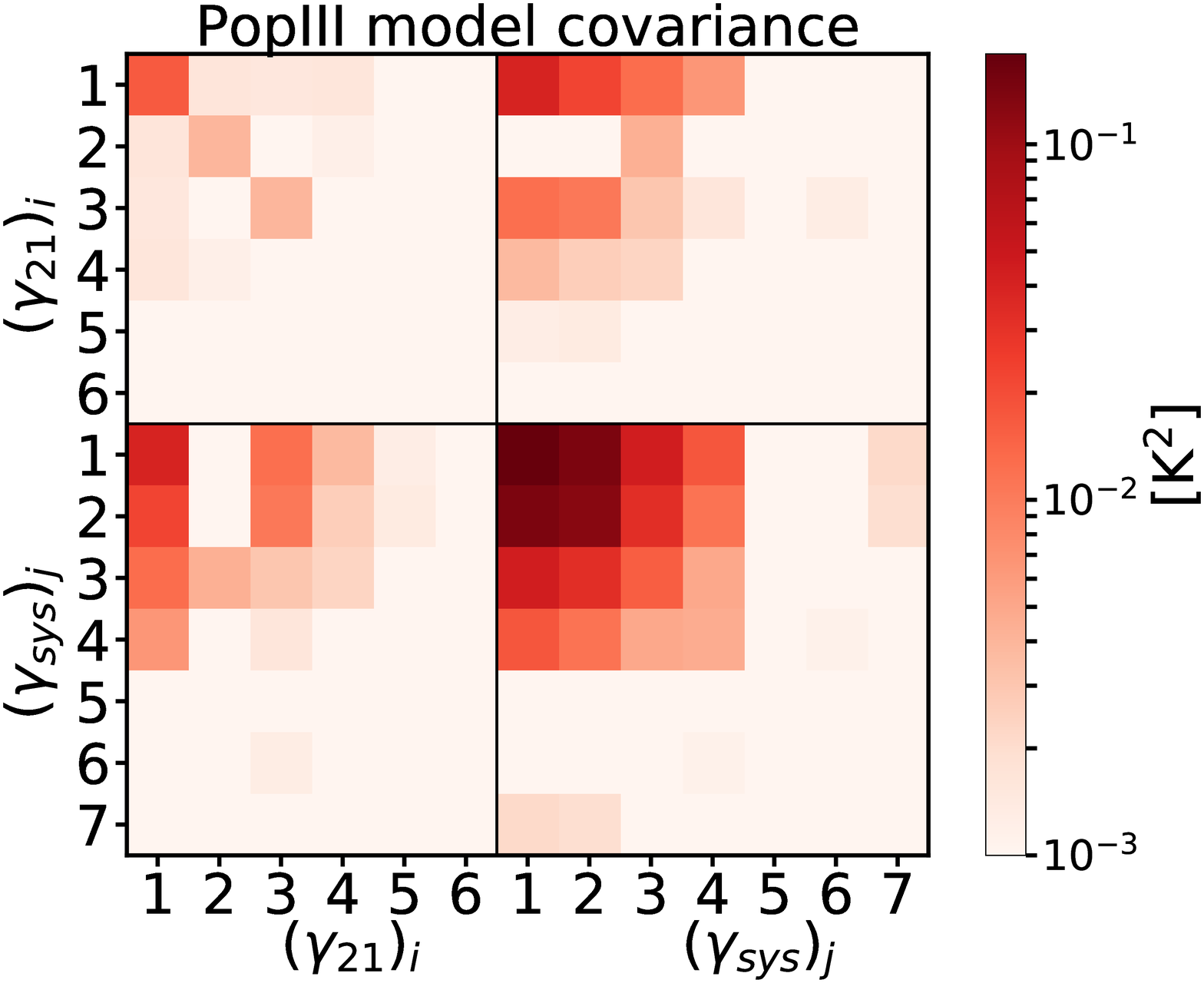}
\caption{{Covariance matrices for the 6 SVD signal parameters, $(\gamma_{21})_i$ for $i\in\{1,2,...,6\}$, and the 7 SVD  systematic parameters, $(\gamma_{sys})_j$ for $j\in\{1,2,...,7\}$, used for fitting the primordial Pop II (left panel) and Pop III (right panel) stellar models. For ease of viewing, the absolute values of the covariances are shown. The vertical and horizontal black lines separate the regions with covariances between signal parameters (top left) and systematic parameters (bottom right). The other two regions are symmetric and show the covariances between parameters multiplying signal and systematic modes.}}
\label{fig:cov}
\end{center}
\end{figure*}

{In this section we demonstrate an essential aspect of our observational strategy: how our data analysis pipeline is able to separate the 21-cm signal from foregrounds measured through a realizable instrument. We model each of these components, signal along with the foreground and instrument systematics, as described below.

In our previous work \cite[][hereafter H12 and H16]{2012MNRAS.419.1070H, 2016MNRAS.455.3829H}, we developed a  foundation for a 21-cm signal extraction pipeline using a Markov Chain Monte Carlo (MCMC) framework. However, we assumed an idealized instrument with exact knowledge of most instrument systematics and the form of the beam-averaged foreground. We have now significantly expanded the initial analyses of H12 and H16 by implementing a robust SVD modeling scheme based upon a pragmatic end-to-end instrument model  (Section~\ref{sec:instrument}). Specifically, the current pipeline (which will be released to the community in a later publication) incorporates the following aspects for the first time:

\begin{itemize}
\item{Full simulations of the antenna beam-weighted foreground. These simulations are based upon beam patterns calculated by the CST electromagnetic simulation package\footnote{https://www.cst.com/} and our diffuse foreground model, described in Section \ref{sec:foregrounds}.}

\item{A calibration model, based upon expected lab measurements and uncertainties, that includes all parameters in Equation~\ref{eq:instresponse}. The instrument model described in H12 included only the antenna reflection coefficient.}

\item{A modeling scheme, detailed below, based upon the implementation of SVD on well-characterized training sets for both the signal and a complete set of instrument and foreground systematics. The SVD technique independently determines the main modes of variation in the signal and systematics. The MCMC algorithm then simultaneously fits all the coefficients associated with the SVD modes to extract the signal. This is a major improvement over our previous use of polynomials (Fourier series) to fit the foreground (reflection coefficient).}
\end{itemize}

The MCMC algorithm in the pipeline samples the likelihood function
\begin{equation}
\ln L(\gamma) = -\frac{1}{2}\sum_{r=1}^{N_r}\sum_{i=1}^{N_\nu} \left[\frac{T_{A,D}^{(r)}(\nu_i)-T_{A,M}^{(r)}(\nu_i,\gamma)}{\sigma_r(\nu_i)} \right]^2,
\label{eq:likelihood}
\end{equation}
where $T_{A,D}^{(r)} (\nu)$ and $T_{A,M}^{(r)} (\nu)$ are the antenna temperature spectra for the data (D) and the model (M), $r$ and $\nu$ indicate the sky direction and frequency channel, respectively, and $\sigma_r(\nu)=T_{A,D}^{(r)}(\nu)/\sqrt{\Delta\nu\Delta t}$ is the thermal noise level in the data for a given frequency bin of width $\Delta\nu$ centered on $\nu$ integrated over a time $\Delta t$. 

We model $T_{A,M}^{(r)} (\nu)$ as a linear combination of (dimensionless) principal modes derived from SVD analyses \citep{2014ApJ...793..102S, 2013MNRAS.433..639P, 2014MNRAS.437.1056V},
\begin{equation}
{T_{A,M}^{(r)}(\nu,\gamma) =  \sum_{i=1}^n(\gamma_{21})_if_i(\nu)+\sum_{j=1}^m(\gamma_{sys})_j^{(r)}{g_j(\nu)\,,}} 
\label{eq:svd}
\end{equation}
where $f_i(\nu)$ and $g_j(\nu)$ are the SVD signal and systematic modes, respectively, and $(\gamma_{21})_i$ and $(\gamma_{sys})_j$ (both with units K) are the coefficients associated with each of them. We fit the entire parameter space, $\gamma=[\gamma_{21},\gamma_{sys}]$, simultaneously \citep[using the \textsc{emcee} code;][]{ForemanMackey2013} in order to account for the covariance between all parameters and ensure self-consistency. This MCMC calculation efficiently and robustly obtains the full posterior distribution.

In this work, we utilize $n=6$ (signal) and $m=7$ (systematic) SVD modes because they are able to fit our fiducial models to within the thermal noise level achieved through 800 hours of integration. For future analyses, however, we are developing a novel technique that will choose the optimal number of modes to use in the pipeline. The details of this key advancement will be described in forthcoming works (Tauscher et al., in prep.; Rapetti et al., in prep.). 

The systematic modes $g_j(\nu)$ are derived from 10,000 simulated datasets which vary the foreground and instrument within expected uncertainties.  This process utilizes Equation~\ref{eq:instresponse}, its inverse, and the fiducial values of the calibration and beam-weighted foreground parameters (Tauscher et al. in prep.). Currently, the signal modes $f_i(\nu)$ are derived from input training sets of 21-cm spectrum simulations (15,000 and 960, respectively) based on two well-motivated ranges of physical models (primordial Pop II and Pop III stars; see Section~\ref{sec:theory}).\footnote{{Each set of simulations was derived by randomly sampling the parameter space surveyed in \citet{2017MNRAS.464.1365M}, with the addition of two parameters that describe the UV and X-ray photon production efficiency in minihalos (i.e., those with $T_{\mathrm{virial}} < 10^4$ K). The Pop III models include only those with Region D extrema in emission.}} In  future work, the signal models will be  combined into a single training set.

\begin{figure*}[t!]
\begin{center}
\includegraphics[width=0.8\textwidth]{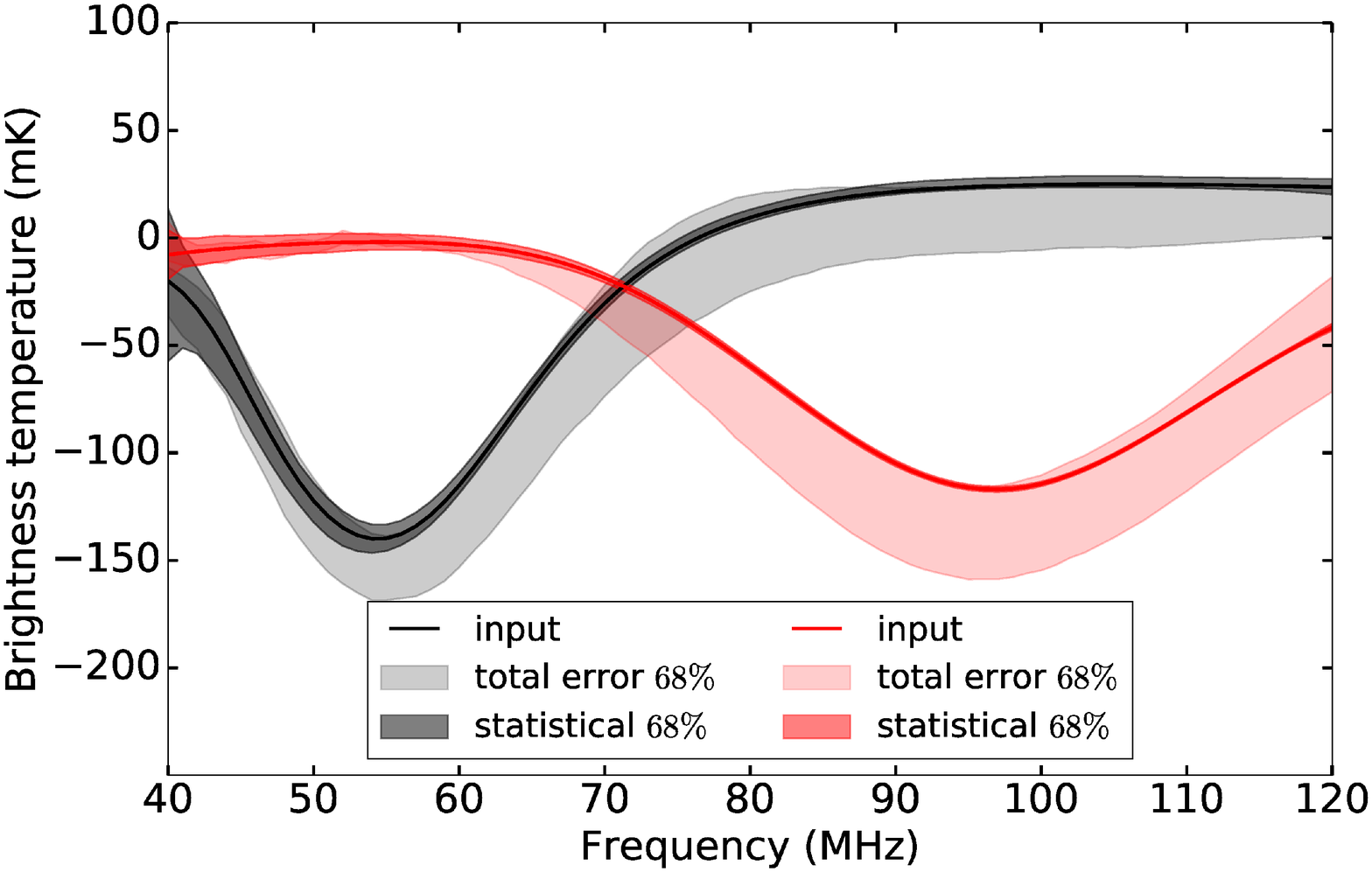}
\caption{The extracted 21-cm spectra with 68\% confidence intervals for models with primordial Pop II (red) and Pop III (black) stars expected using the DARE instrument parameters and 800 hours of observation. The dark bands represent thermal (statistical) noise from the sky. The total uncertainty, including statistical plus systematic effects from the instrument and foreground, is shown by the lighter bands, which are dominated by the covariance between the SVD signal and systematic modes.}
\label{fig:finalspectra}
\end{center}
\end{figure*}

The Bayesian nature of the MCMC permits the incorporation of key prior knowledge on the instrument calibration and foregrounds when retrieving the posterior probability distribution of the model parameters. In the instrument simulations, we account for all the identified uncertainties and priors, including a 50 mK constraint on the beam-averaged foregrounds from measurements of the induced polarization. Even though, at this stage, the induced polarization is used only as a prior on the antenna temperature, $T_A$, in future work, all four Stokes parameters will be included in the likelihood function.}

{Figure~\ref{fig:modes} shows the SVD modes used in this work. The left and middle panels contain the signal modes for the models of primordial Pop II and Pop III stars, respectively. For the purpose of reducing the covariance between our parameters, $\gamma$, it is important that the SVD systematic modes  are as orthogonal as possible (i.e. have a minimal dot product) with the SVD signal modes. When comparing one signal mode with the systematic mode of the same order (color) in Figure~\ref{fig:modes}, we note that the shapes of the modes are sufficiently different to enable a clean extraction of the signal (see Figure~\ref{fig:finalspectra}).}

{Figure~\ref{fig:cov} shows the covariance matrix of the 6 signal parameters, $(\gamma_{21})_i$ with $i\in\{1,2,...,6\}$, and the 7 systematic parameters, $(\gamma_{sys})_j$ with $j\in\{1,2,...,7\}$. The top left corners within each of the 4 regions in both panels of Figure~\ref{fig:cov} demonstrate that the lowest order signal modes have enough similarities in shape with the first 3-4 systematic modes to generate only modest covariances.

By simultaneously fitting all parameters, $\gamma$, and marginalizing over the systematic parameters, $\gamma_{sys}$, we are able to clearly separate the signal from the systematics despite those covariances, as shown in Figure~\ref{fig:finalspectra}. The  widths of the uncertainty bands result from the covariances between the signal parameters, which depend on the level of overlap between the signal and systematic modes. This will be explored in detail in upcoming work (Tauscher et al., in prep.).}

In summary, for the DARE instrument parameters discussed in Section~\ref{sec:instrument} and 800 hours of total integration above the lunar farside, our signal extraction pipeline {recovers the spectra} and uncertainties for two representative models (Pop II and Pop III star models) shown in Figure \ref{fig:finalspectra}. In addition to the 21-cm signal, the pipeline simultaneously fits the SVD modes of the receiver, beam, and foreground utilizing prior information and on-orbit measurements.  With an average RMS of $\approx$17 mK, we recover the major features in the spectra and differentiate between different stellar population models.

\begin{figure*}[tb]
\begin{center}
\includegraphics[width=0.47\textwidth]{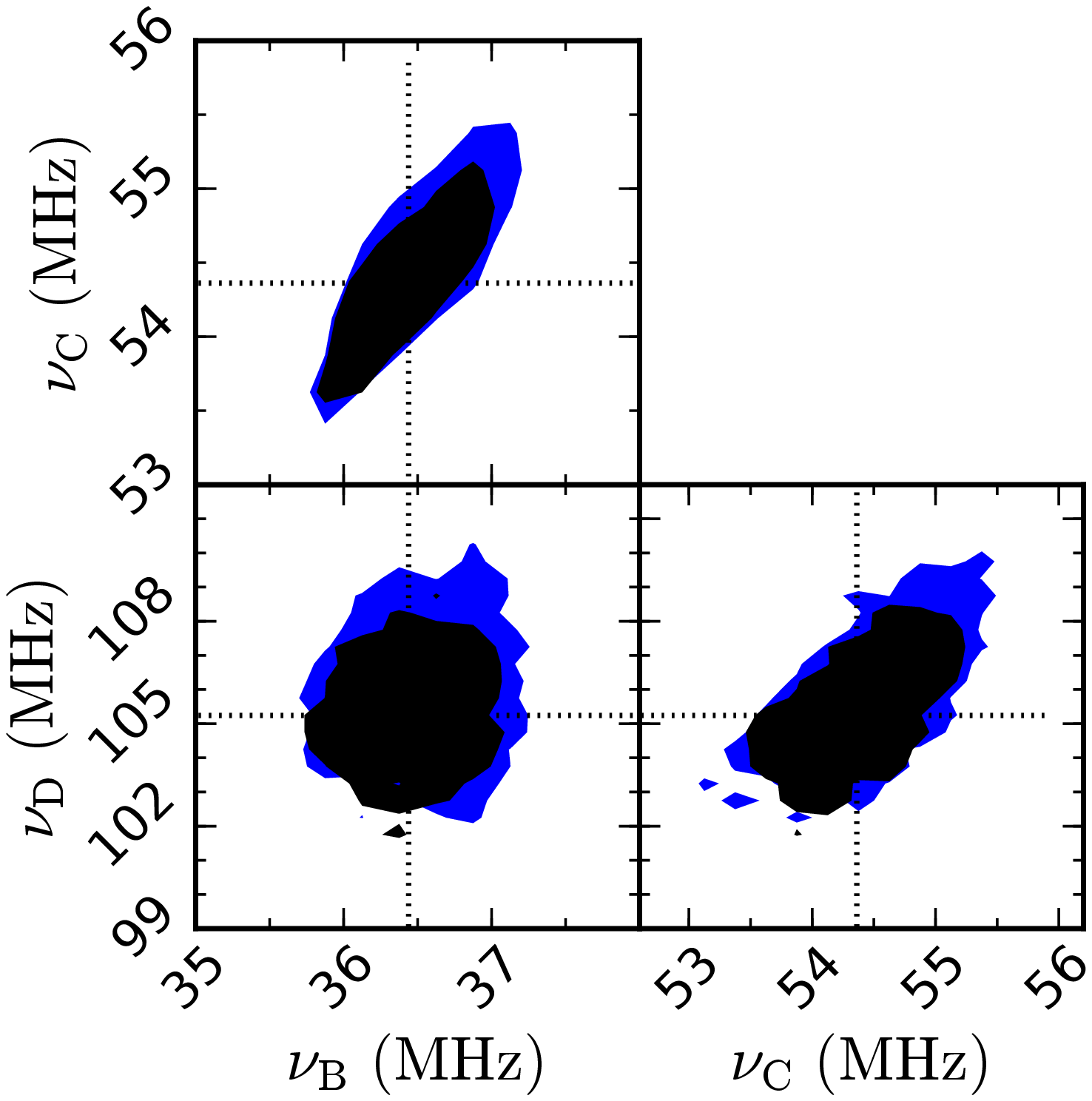}\hfil%
\includegraphics[width=0.47\textwidth]{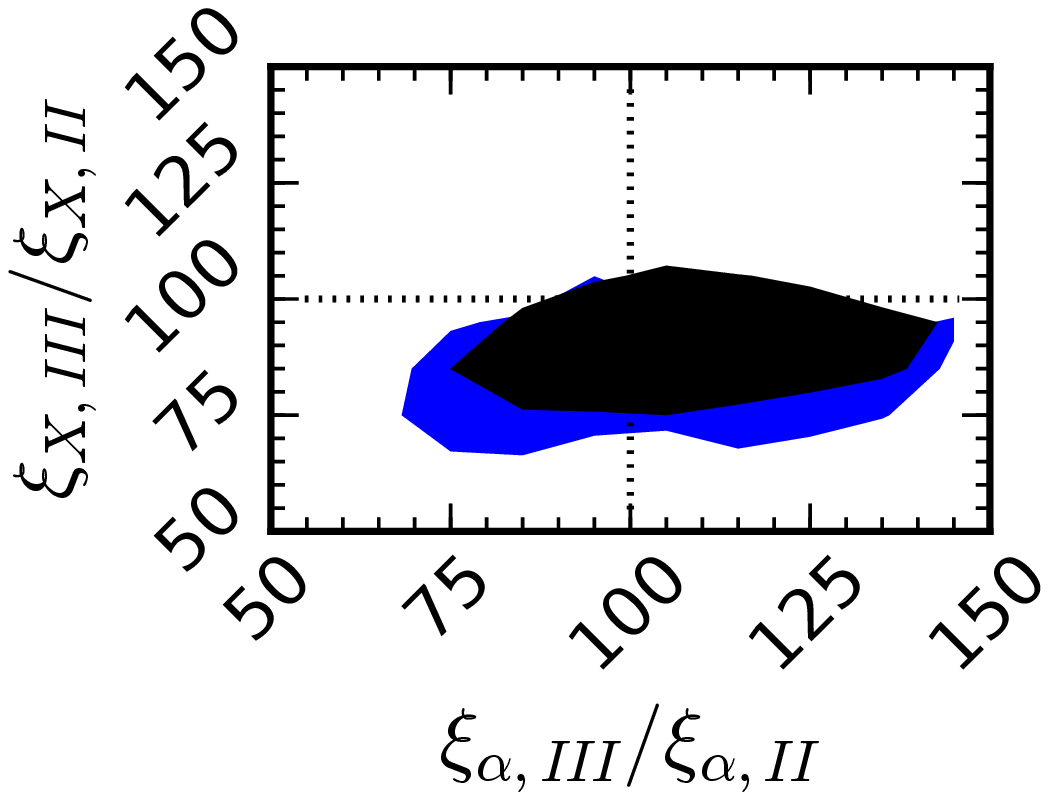}
\caption{The panels illustrate examples of constraints on the global 21-cm extrema frequencies (left), UV photon production efficiency ($\xi_{\alpha}$) and X-ray heating efficiency ($\xi_X$) between models with Pop III and Pop II stars (right) using the calibrated 21-cm spectrum.  Dotted black lines indicate the ``true'' input values.  The contours are at the 68\% confidence intervals {\bf using 23 (black) and 30 (blue) mK average RMS uncertainties over the observed band}.}
\label{fig:physicalparameters}
\end{center}
\end{figure*}

\section{Physical Parameter Estimation}
\label{sec:estimates}

With the calibrated spectra and uncertainties in Figure \ref{fig:finalspectra}, it is straight-forward to estimate when the first luminous objects ignited and began reionizing the Universe.  Since redshift maps directly to frequency, measurements of the extrema frequencies from the 21-cm spectrum determine when major events occurred in a mostly model-independent fashion \citep{2016MNRAS.455.3829H}.  The frequency of the Region B extremum ($\nu_B$) determines the $z$ at which the UV background activates the 21-cm transition (i.e., first stars ignition).  This clean and accurate measurement delineates the nature of the first stars, especially considering that no observational bounds currently exist.  Using a Pop III model as a working example, DARE will extract $\nu_B$ with a $1\%$ (0.4 MHz) uncertainty (68\% confidence).  Similarly, the redshift when the first black holes began accretion is measured from the Region C extremum frequency ($\nu_C$) with 1\% (0.6 MHz) uncertainty.  The redshift of the beginning of reionization is measured from the extremum $\nu_D$ with 2\% (2 MHz) uncertainty. Different models (Figure \ref{fig:global_signal}) yield similar uncertainties for the extrema frequencies.\footnote{{Note that the extrema locations are determined from the full signal model on each step of the MCMC, i.e., these quantities have not assumed a cubic spline form for the signal as in some previous works \citep{2012MNRAS.419.1070H}.}}

The characteristics of the first stars and galaxies, along with the history of the early Universe, are determined from modeling of the calibrated spectrum.  First, the history of reionization in the early Universe is characterized by the {\HI} fraction ($x_{\it{\HI}}$) and the IGM kinetic temperature ($T_K$) at $z\sim11$.  Our modeling of features in Region D using DARE's sensitivity yields uncertainties of 5\% and 10\% for $x_{\it{\HI}}$ and $T_K$, respectively.

Next, the features in the 21-cm spectrum at the lowest frequencies depend upon the stellar populations that dominate the UV background; if, for example, Pop III star formation is efficient, we should expect features of the signal to occur at lower frequencies (higher redshift) than if Pop II stars dominate the background because Pop II stars form in more massive halos which do not become abundant until relatively late times (low redshift). DARE's sensitivity can separate the effects of the broad classes of Pop II and Pop III stellar models considered in this work (see Figures \ref{fig:global_signal} and \ref{fig:finalspectra}), subject to the assumed calibration of the Pop II contribution \citep[see Section~\ref{sec:theory};][]{2017MNRAS.464.1365M} and the model for Pop III stars. A useful metric for gauging the influence of Pop III is the ratio of UV production efficiencies for Pop III compared to Pop II stars, $\xi_\alpha \equiv \xi_{\alpha,III}/\xi_{\alpha,II}$. {The value of $\xi_{\alpha,II}$ is drawn from the BPASS models \citep{2009MNRAS.400.1019E} assuming solar metallicity, while $\xi_{\alpha,III}$ is allowed to vary freely. DARE constrains $\xi_{\alpha,III}$} with 25\% uncertainty in Figure \ref{fig:physicalparameters}.

The characteristics of the first X-ray sources (Region C in Figure \ref{fig:global_signal}) are inferred from the ratio of X-ray heating efficiencies between Pop III and Pop II stars. {Analogous to the UV constraints, the Pop III X-ray efficiency, $\xi_{X,III}$, is allowed to vary freely, while $\xi_{X,II}$ is anchored to the local relation between X-ray luminosity and SFR \citep{Mineo2012a} assuming high-mass X-ray binaries are the dominant source. DARE can measure $\xi_{X,III}$} with 15\% uncertainty (see Figure \ref{fig:physicalparameters}).  Further modeling plus multi-wavelength observations \citep[e.g., the cosmic X-ray background;][]{2017MNRAS.464.3498F} may help to better constrain the identity of the Universe's first X-ray sources, whether they be black hole X-ray binaries, hot gas in star-forming galaxies, or proto-quasars.

Finally, before concluding, we emphasize that these Pop III models are quite simple, as, for example, they neglect an explicit treatment of feedback. As a result, the interpretation of the precise value of $\xi_{\alpha,III}/\xi_{\alpha,II}$ may be considerably more complex in practice, but the finding that both values are non-zero is robust. 

\section{Concluding Remarks}
\label{sec:conclusion}

To achieve the science potential of 21-cm global spectral observations, we proposed an observational strategy that carefully considers  the local environment, the instrument, and the methods for signal extraction.  A lunar-orbiting experiment above the Moon's farside has the best probability of measuring the 21-cm spectrum since this environ is free of ionospheric effects and human-generated radio frequency interference.  

Signal extraction in the presence of bright foregrounds is the greatest challenge for all observations of the 21-cm cosmological spectrum. Utilizing Singular Value Decomposition to model the foreground and instrument along with a Markov Chain Monte Carlo numerical inference technique to survey parameter space, we showed that it is possible to accurately recover the expected features in the spectrum in the presence of bright foregrounds with the  instrument characteristics of the {\it Dark Ages Radio Explorer} (DARE) for $\approx$800 hrs of integration. To separate the signal from the foreground, the antenna system  must be well-characterized requiring temperature control and precise beam directivity measurements on the ground and in-space.  In addition, a model-independent constraint on the foreground from polarimetric observations is an important element in the signal extraction.

From the extracted 21-cm spectrum (including confidence intervals), we showed that meaningful constraints can be placed upon the physical parameters of primordial radiating objects.  The redshift for the commencement of first star formation and X-ray emission from the first accreting black holes along with the redshift of the beginning of reionization can be inferred to within $\approx$ a few percent.  The 21-cm signal is also uniquely sensitive to the different radiation effects produced by Pop II and Pop III stellar models.  Specifically, the UV production and X-ray heating efficiencies can be constrained, thus determining which stellar population was dominant within the first galaxies.  Finally, the history of reionization in the early Universe can be characterized by the redshift evolution of the {\HI} ionization fraction ($x_{\HI}$) inferred from the 21-cm spectrum.

Accurate parameter estimation is a core capability required for 21-cm global signal observations and interpretations.  Bayesian methods have significant potential for 21-cm observations \citep{2016MNRAS.455.4295G, 2016MNRAS.461.2847B}. They have proven to be successful {for other experiments targeting weak signals}, including CMB observations \citep{2016A&A...594A..13P, 2016A&A...596A.108P} and the LIGO gravitational wave detections \citep{2016PhRvL.116f1102A, 2015PhRvD..91d2003V}.  The next step in the analyses of the global 21-cm spectrum is to construct a likelihood function allowing differentiation between differing physical models for the first halos. Similarly,  modeling different levels of structure in the beam-convolved foregrounds needs a refined Bayesian approach. This is a highly computationally intensive process. {We are refining and extending our SVD modeling approach towards these goals. In addition,} recent developments of Nested Sampling algorithms for high dimensional parameter spaces which operate in massively parallel computer architectures \citep{2015MNRAS.450L..61H} have great potential for 21-cm cosmology applications. 

In conclusion, measurements of spectral features in the 21-cm spectrum will answer key science questions from the NRC Astrophysics Decadal Survey: ``What were the first objects to light up the Universe and when did they do it?"  With a clean measurement within the radio-quiet environs of the Moon's farside and proven technology, the 21-cm global signal will open a new window of discovery into the early Universe.

\acknowledgments{This research was supported by the NASA Ames Research Center via grants NNA09DB30A, NNX15AD20A,  NNX16AF59G, as well as by a NASA SSERVI Cooperative Agreement to JB. The theoretical work was partially funded by NASA ATP grant NNX15AK80G to SF and JB. DR is supported by a NASA Postdoctoral Program Senior Fellowship at the NASA Ames Research Center, administered by the Universities Space Research Association under contract with NASA. We thank the following colleagues for their important contributions including Jill Bauman, Jay Bookbinder, Matthew D'Ortenzio,  Robert Hanel, Butler Hine, Joseph Lazio, Stephanie Morse, Ken Galal, Laura Plice, Gary Rait, Jeremy Stober, and Eugene Tu.  We also benefited from helpful input from James Aguirre, Mina Cappuccio, James Condon, Steven Ellingson, Bryna Hazelton, Danny Jacobs, Mark LaPole, Adam Lidz, John Mather, Harry Partridge,  Aaron Parsons, Jonathan Pober, Greg Schmidt, Greg Taylor, and Sandy Weinreb.} 
\bibliography{DARE}
\end{document}